\documentclass[12pt]{article}
\raggedright

\begin{document}

\title{Realism in the Realized Popper's Experiment}

\author{
Geoffrey Hunter\\
{\small\sl Department of Chemistry},\\
{\small\sl York University, Toronto, Canada  M3J 1P3}}

\maketitle

\vspace*{-0.25in}

\begin{abstract}
                  The realization of Karl Popper's experiment by 
Shih and Kim (published 1999) produced the result that Popper hoped for: no 
``action at a distance'' on one photon of an entangled pair when a 
measurement is made on the other photon.  This experimental result 
is interpretable in local realistic terms: each photon has a definite 
position and transverse momentum most of the time; the position measurement on 
one photon (localization in a slit) disturbs the transverse momentum of that 
photon in a non-predictable way in accordance with the uncertainty 
principle; however, there is no effect on the other photon (the photon that is not in a slit)
no action at a distance.  The position measurement (localization within a slit) of the one 
photon destroys the coherence (entanglement) between the 
photons;~i.e.~decoherence occurs.  This can be understood physically 
as an electromagnetic interaction between the photon in the slit 
and the electrons of the atoms in the surface of the solid that 
forms the slit.  

For an individual entangled pair, the detected position of the other (not in a slit) photon 
is the data for the calculation of the transverse momenta 
of both photons of this pair; for the photon interacting with the slit this calculation produces
its momentum as it enters the slit (just before the interaction takes place), but its momentum 
after the localization (as it emerges from the slit) is not easily calculable 
because it depends upon the structure of the electromagnetic field of 
the photon, upon how this interacts with the atomic electrons in the 
surface of the slit, and upon the impact parameter of the incident 
photon with the centerline of the slit; this impact parameter is different 
for every incident photon, being a statistical aspect of the beam of 
photons produced by the experimental arrangement.   This complex physical interaction 
is subsumed in the statistics of the uncertainty principle.

This realistic (albeit retrodictive) interpretation of the Shih-Kim realization of what Popper called his
``crucial experiment'' is in 
accord with Bohr's original concept of the nature of the uncertainty 
principle, as being an inevitable effect of the disturbance of the 
measured system by the measuring apparatus.  These experimental results are also in accord with the 
proposition of Einstein, Podolski and Rosen's 1935 paper: that quantum mechanics provides only a
statistical, physically incomplete, theory of microscopic physical processes.
\end{abstract}

\newpage

\section{Popper's Concept of the Crucial Experiment}
Karl Popper was a philosopher who was deeply concerned with the interpretation of quantum
mechanics since its inception in 1925.   In a book originally published in 1956 
he proposed an experiment which he described as {``an extension of the Einstein-Podolsky-Rosen
argument''} \cite[pp.ix,27-30]{Popper}.  In summary: two photons are emitted simultaneously from 
a source that is mid-way 
between (and colinear with) two slits A and B; the coincident diffraction pattern is observed beyond 
each slit twice: once with both slits present, and again with one slit (slit B) 
``wide open''.\footnote{T.~Angelides has restated this as ``slit B being absent''.}

\hspace*{0.25in}Popper (in common with Einstein, Podolsky and Rosen and with the proponents of the Copenhagen
interpretation of quantum mechanics) believed that quantum theory predicts that localization of
a photon within slit A will not only cause it to diffract in accordance with Heisenberg's uncertainty
relation, but will also cause the other particle at the location of slit B to diffract by the 
same angle - regardless of whether slit B is present or not.
Such spatially separated correlations would violate the causality principle of Special Relativity; i.e.~that
physical interactions cannot travel faster than the speed of light; thus Einstein called them
``spooky actions at a distance'' \cite{Pais}.

\hspace*{0.25in}Popper was ``inclined to predict'' \cite[p.29]{Popper} that in an actual experimental test, the photon at
the location of the absent slit B would not diffract (``scatter'') when its partner photon (with which it
is in an entangled state) is localized within slit A.  He emphasized that ``this does not mean that quantum
mechanics (say, Schr\"odinger's formalism) is undermined''; only that ``Heisenberg's claim is undermined 
that his formul\ae\ are applicable to all kinds of indirect measurements''.

\hspace*{0.25in}Popper \cite[p.62]{Popper} notes that Heisenberg agreed that {\em retrodictive} values of the position
{\em and} the momentum can be known by knowing the position of a particle (e.g.~as it passes through
a small slit) followed by a measurement of the momentum of the particle after it has passed through 
the slit - from the position in the detection plane where it is located.   Feynman has also emphasized that
the uncertainty principle does not exclude  {\em retrodictive} inferences about the simultaneous 
position and momentum of a particle\\ \cite[Vol.3,Ch.2,pp.2-3]{Feynman}, but  the uncertainty
principle does exclude precise {\em prediction of both} position or momentum.
This view that retrodictions of classical particle trajectories\footnote{i.e.~positions and momenta of 
the particle everywhere along its path.} are consistent with quantum
mechanics, is inconsistent with Bohm's interpretation (Bohmian mechanics) in which the particle 
follows a non-classical trajectory under the influence of the ``quantum potential'' \cite{Holland}.  

\hspace*{0.25in}Popper also notes  
 that momentum values are usually inferred from two sequential position measurements; in this regard
he concurs with  Heisenberg, who wrote \cite[p.62]{Popper}:
\begin{quote}
``The \ldots most fundamental method of measuring velocity [or momentum] depends on the 
determination of position at two different times \ldots it is possible to determine {\em with any 
desired degree of accuracy} the velocity [or momentum] before the second measurement was made''
\end{quote}
The whole of Popper's rebuttal of what he calls ``the great quantum muddle'' is 
very inciteful \cite[pp.50-64]{Popper}:
he emphasizes that while {\em predictions of future events} (such the trajectory of an individual particle)
are only statistical with statistics limited by the Uncertainty Principle, that {\em precise 
 retrodiction of past events} (the positions and momenta of the particle along its trajectory) are not 
only possible - ``they are {\em tests} of the theory''\\ \cite[p.63,last line]{Popper}.  Indeed precise
inferences of particle positions and momenta are widely used in the analysis of  observations in
high-energy particle accelerators, by which the modern plethora of ``fundamental particles'' have been
discovered.  This is emphasized by another quote from Popper  \cite[p.39]{Popper}:
\begin{quote}
``my assertion [is] that most physicists who honestly believe in the \\\ \ Copenhagen interpretation
do not pay any attention to it in actual practice''.
\end{quote}

\subsection{Measurements: Selections and Interactions}

Popper distinguished between two types of 
measurements on particles: 
\begin{description}
\item{Selective Measurements} simply select particles from a larger ensemble of particles;~e.g.~a 
slit selects those particles in a beam (larger than the slit) that happen to collide with the selecting screen 
{\em within} the slit, rather than on either side of it (when they will be absorbed or reflected).
\item{Interactive Measurements} disturb the physical state of the particle; e.g.~when a particle is 
absorbed within the emulsion of a photographic detecting plate, or within a photo-multiplier detector.
\end{description}
However, selection (location of a particle's position) by a slit, may subsequently (after the moment of
selection) lead to an interaction with the slit that causes the momentum of the particle to change in
an unpredictable way, because the precise trajectory of the particle as it passes through, and interacts
with the slit, cannot be known in principle; determination of its trajectory would inevitably
involve disturbing its trajectory; this is the essence of the Uncertainty Principle interpreted as
a disturbance of the particle in which the  action\footnote{``action'' in the physical sense as
momentum$\times$displacement or energy$\times$time} involved cannot be smaller than 
Planck's constant - the quantum of action.

\hspace*{0.25in}Reflections at a mirror do change the direction of the momentum of a photon in a 
predictable way, 
but they do not change the magnitude of its momentum.  Whether reflection at a mirror should be
regarded as a measurement is a moot point.  Selective reflection or transmission in a beam
splitter is also in this category of  apparently  non-interactive changes in momentum.
The deflections of an electron in a Stern-Gerlach apparatus \cite[pp.404-416]{Holland}
also appear to be a momentum-changing, but otherwise non-interactive ``measurement''.

\hspace*{0.25in}Studies by Bell and others \cite{BellsBook}, have concluded that any realistic
interpretation of quantum mechanics must involve an essential non-locality of physical interactions.
In this regard it is salutory to reflect that Planck's constant is a constant of 
Action, and that the nature of action (momentum$\times$distance, energy$\times$times) is 
intrinsically non-local (regardless of quantization) because one of its factors (distance or time) 
is an extension in space or time.  Likewise angular momentum, with the same physical dimensions
as action, being an attribute of rotational motion 
is classically a property of an extended object.
\filbreak

\section{The Real Experiment of Shih and Kim}

	Yoon-Ho Kim and Yanhua Shih carried out a modern realization of Popper's experiment;
their report was published almost simultaneously three times 
\cite{Shih1,Shih2,Shih3}.\footnote{The triple publication may have occured because the paper was 
thought to be of such fundamental importance that it should reach the widest possible audience, 
or perhaps it was a result of simultaneous submissions in anticipation of the 
predisposition of journal editors and their referees against accepting an experimental result that 
is in disagreement with the prevailing beliefs of the majority of physicists.}
Other experiments that don't confirm the
widely held interpretation of quantum mechanics have also 
been reported \cite{Bell1972}; in his book
John Bell wrote \cite[p.60]{BellsBook}:
\begin{quote}  
``Of course, any such disagreement, if confirmed, is of the utmost importance''
\end{quote}  

\subsection{Quotes from Kim and Shih's Paper}

These quotations are drawn from \cite[\S 1,pp.1849-1950]{Shih1}:
\begin{itemize}
\item Karl Popper believed that:
\begin{quote}
``the quantum formalism {\em could} and {\em should} be interpreted \\
\ \ \ realistically \ldots the same view
as Einstein''.
\end{quote}
\item Popper's thought experiment was designed to show  
\begin{quote}
``that a particle can have both precise position \\\ \ \ and momentum at the same time''.
\end{quote}
\item The experiment is 
\begin{quote}
``a correlation measurement of an entangled two-particle system''.
\end{quote}
\item Popper's thought experiment is strikingly similar to the Einstein, Podolski, Rosen {\em gedanken}
experiment (EPR) proposed in 1935 \cite{EPR}; Popper also thought of it ``in the early 1930s'' 
\cite[p.1849]{Shih1}, but unlike EPR (which became part of the folklore of physics) Popper's 
{\em gedanken} experiment was remembered by only a few physicists, perhaps because it wasn't
published in a scientific periodical.
\item In the realization of Popper's experiment
\begin{quote}
``{\bf it is astonishing to see that the experimental \\\ \ \ results agree with Popper's prediction}''.
\end{quote}
[emphasis added].
\end{itemize}

\subsection{Experimental Parameters}\label{exptdata}
\begin{itemize}
\item the source is a CW\footnote{CW = continuous wave} argon-ion laser producing highly 
monochromatic ultra-violet radiation of wavelength, $\lambda= 351.1$ nm.
\item a pair of coherent (i.e.~entangled) visible photons are produced by Spontaneous Parametric 
Down-Conversion (SPDC) in a BBO crystal.\footnote{BBO = $\beta$ barium borate}
\item the length of the BBO crystal $= 3$ mm.
\item the diameter of the photon beam $= 3$ mm.
\item the wavelength of both of the entangled photons $= 702.2$ nanometers (nm).
\item the two photons emerge from the BBO crystal in the same direction; they are separated
by a polarizing beam-splitter, which directs them in different directions.
\item the width of slit A and of slit B (when present) $= 0.16$ mm.
\item the diameter of the avalanche photodiode detectors (D1 and D2) $= 0.18$ mm.
\item the distance from the SPDC source (the BBO crystal) to slit A  was 1255 mm.
\item the distance (optical path length) from the SPDC source to slit B was 1245 mm.
\item the difference of 10 mm between these path lengths may be due to the presence of the lens LS
in the path to slit A; the two optical path lengths (from the BBO crystal to D1 and to D2) are 
presumably the same in order to achieve the coincident detection that is an essential feature
of the measurements.
\item the distance (optical path length) from slit B to detector D2 was 500 mm.
\item During the measurements, D1 was in a fixed position (\underline{not} scanned along the $y$-axis) 
close to a collection lens (of 25 mm focal length) with the lens close to slit A; this lens was
designed to direct \underline{every} photon passing through slit A into detector D1.
\item a lens (focal length=$f$=500 mm) is placed in the path of the photon going towards slit A and
detector D1; slit A is $2f$ from the lens, and the plane of slit B (regardless of whether slit B is
actually present) is also $2f$ from the lens as measured from the lens, back along the path of
the slit A photon to the BBO crystal, and then forwards from the BBO crystal along the path of
the photon going towards slit B.  

In their report \cite[Fig.4,p.1854]{Shih1} they say that this makes
a ``ghost image'' of slit A at slit B (regardless of whether slit B is present or ``wide open'').  
The significance of this is not clearly explained, although it was presumably essential to the 
successful execution of the experiment; they do, however, cite a previous
publication as containing an explanation of the principle of the ghost image \cite{Shih4}.

   On the other hand,
Angelides claims that:\footnote{Personal communication by email, April 2002.}
\begin{quote}
The presence of the lens (LS) before slit A is \\\ redundant, despite the Kim-Shih claim that: \\
\ \ ``The use of LS is to achieve a `ghost image' of slit A at screen B''.
\end{quote}
It is true that if light were being emitted through slit A towards the lens LS, and then reflected in the
beam-splitter back to the BBO crystal, and then reflected out of the crystal back to the beam
splitter, and then onwards towards screen (slit) B, that the real image of slit A at screen B would be
the same size as slit A, because slit A and screen B are equidistant from the lens LS at twice its
focal length.  However, since in fact light is not radiating in this direction in the actual experiment,
the significance of the ``ghost image'' achieved by the lens LS is unclear and controversial;
it is not involved in the subsequent analysis here. 
\end{itemize}

\subsection{Classical Theory of Diffraction}\label{classtheory}

The relative light intensity beyond a single slit in a screen is given from 
the classical theory of diffraction \cite[pp.214-216]{Longhurst} by: 
\begin{eqnarray}\label{diffint}
{\rm relative\ intensity} = 
\left[\frac{\sin\!\left(\frac{\pi\,s\,y}{\lambda\,D}\right)}{\left(\frac{\pi\,s\,y}{\lambda\,D}\right)}\right]^2
\end{eqnarray}
where $s$ is the slit width, $\lambda$ is the wavelength of the light, $D$ is the distance of the 
observing plane\footnote{the scanning plane of photon-detector D2 in this experiment} from 
the screen containing the 
slit,\footnote{The observing plane and the slit-screen are parallel; they are perpendicular
to the incident light beam.} and $y$ is the displacement in the observing plane
from the axis of the incident beam (the $x$-axis).  
Graphs of the peaks and troughs generated by (\ref{diffint}) are given in \cite[pp.215,223]{Longhurst}.
The central maximum (relative intensity = 1) occurs when 
$y=0$.\footnote{$y=0$ is the point in the observing plane where the undiffracted photons of the 
 incident beam meet it.}  

\subsubsection*{The first minimum (zero intensity) occurs when:}
\begin{eqnarray}\label{firstmin}
\left(\frac{\pi\,s\,y}{\lambda\,D}\right) = \pi \quad \Rightarrow \quad y = \frac{\lambda\,D}{s}
\end{eqnarray}
 For $\lambda=702.2$ nm, $s=0.16$mm, and $D=500$ mm, this produces:
\begin{eqnarray}\nonumber
y &=& \frac{702.2\times 10^{-9}\times 500\times 10^{-3}}{0.16\times 10^{-3}}\\
   &=& 2.194{\scriptstyle 375}\ {\rm mm}
\end{eqnarray}
which is the value (2.2 mm) evident from the experimental Figure 5 \cite[p.1855]{Shih1}.

\subsubsection{Diffraction in terms of the Uncertainty Relation}\label{diffHUP}

Insertion of the de Broglie relation:
\begin{eqnarray}\nonumber
\lambda = \frac{h}{P}
\end{eqnarray}
into \ref{firstmin} produces a form that resembles the Uncertainty Relation:
\begin{eqnarray}\label{HUPdiff}
 y\, P  = \frac{h\,D}{s}
\end{eqnarray}
where $s$ is the slit width and $P$ is the momentum of a photon diffracted towards
the \underline{first minimum} in the diffraction pattern; $y$ is the displacement from the $x$-axis
of the first minimum on the screen, the screen being at distance $D$ from the plane of the slit.

This formula (\ref{HUPdiff}) has been presented by Feynman as a simple ``derivation'' of the
Uncertainty Relation; in this context (\ref{HUPdiff}) is re-interpreted as:
\begin{eqnarray}\label{HUPdiffdelta}
 \Delta y\, \Delta P_y  = \frac{h\,D}{s}
\end{eqnarray}
in which $\Delta y$ is the ``uncertainty'' in the position of the photon in the $y$ direction, and
$\Delta P_y$ is the ``uncertainty'' in the $y$ component of the photon's momentum.  Thus the
position of the \underline{first minimum} in the diffraction pattern ($y$ in eqn.(\ref{HUPdiff}))
is identified with the spread of a beam of photons after it passes through a slit of width $s$.
This is relevant to the interpretation of the two experimental curves in Figure 5 of \cite{Shih1} (below).

\subsection{Momenta and Uncertainty Products}

	The product of the uncertainties in the position and momentum of the photons is 
calculated as follows:
\subsubsection{Photon Properties} 
 The two photons
are emitted at the same time from a small volume\footnote{From the experimental data 
(\S\ref{exptdata}) the source is a cylinder 3 mm in diameter and 3 mm long.} 
surrounding the point $\{x$$=$$0, y$$=$$0\}$.\footnote{The line connecting the two slits is 
designated the $x$ axis by Shih and Kim.}  
They have a wavelength
of 702.2 nm, and hence the energy of each photon is:
\begin{eqnarray}\nonumber
E= h\,\nu = h\,c/\lambda &=& 6.6261{\scriptstyle 76}\times 10^{-34}
\times2.9979{\scriptstyle 25}\times 10^{8}/(702.2\times 10^{-9}) 
\\
&=& 
2.828{\scriptstyle 93}\times 10^{-19}\;{\rm Joules}
\end{eqnarray}
Thus the momentum of each photon in the direction of propagation is given by:
\begin{eqnarray}\nonumber
P = h\,\nu/c = E/c &=& 2.828{\scriptstyle 93}\times 10^{-19}/2.9979{\scriptstyle 25}\times 10^{8}
\\
&=&
9.436{\scriptstyle 31}\times 10^{-28}\;{\rm Kg.metre/sec}
\end{eqnarray}
and the dynamic mass of each photon is 
\begin{eqnarray}\nonumber
M= P/c &=& 9.436{\scriptstyle 31}\times 10^{-28}/2.9979{\scriptstyle 25}\times 10^{8}
\\
&=&
3.147{\scriptstyle 61}\times 10^{-36}\;{\rm Kg}
\end{eqnarray}
\subsubsection{Transverse Momentum: Slit B Present}
From Kim and Shih's Figure 5, when slit B is present the photons emerging from slit B are deflected
(diffracted) up to about 2.0 mm over a distance (from the slit) of 500 mm; 
a feature of the wider curve of the experimental results \cite[Fig.5,p.1855]{Shih1}
is a diffraction minimum at $y=2.2$ mm with one data point showing the onset of the second 
diffraction maximum; this is predicted by the classical theory of
diffraction (eqn.(\ref{firstmin}) in \S\ref{classtheory} above).  

\hspace*{0.25in}
In view of (\ref{HUPdiff}) it is appropriate to work out the momentum of photons diffracted towards
the \underline{first minimum} in the observed diffraction pattern; i.e.~those diffracted to reach detector D2
at $y=2.2$ mm.   The length of the path from slit B to the first minimum is:
\begin{equation}\label{diagdist}
D' = \sqrt{D^2+y^2} = \sqrt{500^2+2.2^2} = 500.00484 {\rm mm} \approx 500 {\rm mm}
\end{equation}
The time taken for the photon
to travel this distance is $0.500/c=1.66{\scriptstyle 78}\times 10^{-9}$ seconds.  

In this time it travels 2.2 mm in the $y$ direction; i.e.~its speed in the $y$ direction is: 
\begin{equation}\label{vy}
v_y = 0.0022/1.66{\scriptstyle 78}\times 10^{-9} = 1.31{\scriptstyle 91}\times 10^{6} {\rm metre/sec}
\end{equation}
and hence its transverse momentum (in the $y$ direction) is:
\begin{eqnarray}\nonumber
P_y = M\,v_y = 3.147{\scriptstyle 61}\times 10^{-36}\times 1.31{\scriptstyle 91}\times 10^{6} 
&=& 4.15{\scriptstyle 20}\times 10^{-30} \;{\rm Kg.metre/sec}
\end{eqnarray}
Multiplication of this transverse momentum by the slit width of 0.16 mm produces a position-momentum
uncertainty product of:
\begin{eqnarray}\nonumber
\Delta y\,\Delta P_y 
&=& 0.00016\times 4.15{\scriptstyle 20}\times 10^{-30}
\\
&=& 6.6{\scriptstyle 43}\times 10^{-34} \;{\rm Joule.sec}\\
\nonumber
{\rm which\ is\ Planck's\ constant:}\quad h &=& 6.626176\times 10^{-34} \;{\rm Joule.sec}\hfill
\end{eqnarray}
within the two decimal digit precision of the data: i.e.~a slit width of $0.16\pm 0.005$ mm and 
an observed diffraction minimum at $2.2\pm 0.05$ mm.
This result exemplifies Feynman's derivation of the uncertainty principle given above in \S\ref{diffHUP};
specifically equation (\ref{HUPdiffdelta}).

\paragraph{It is noteworthy that Kim and Shih report}\cite[top of p.1856]{Shih1}
\begin{quote}
``the {\em single} detector counting rate of D2 is basically the same as that of the coincidence
counts except for a higher counting rate''.
\end{quote}
In other words: the outer (wider) diffraction peak\footnote{the curve for slit B present} of Figure 5 of 
\cite{Shih1} is obtained regardless of whether the coincidence detection circuit is active or not.
This observation indicates that this peak is produced by the diffraction of the photons incident
upon slit B {\em regardless} of their entanglement with the photons incident upon slit A; 
i.e~it is a single-photon phenomenon; indeed it is nothing more than the central diffraction maximum 
predicted by the classical theory of light \cite[p.214]{Longhurst}.

\hspace*{.25in}The higher counting rate is explained by the effective size of the source only being about
0.16 mm diameter for coincidence counting, whereas for non-coincidence counting it is the full diameter
of the laser beam of 3 mm.

\subsubsection{Transverse Momentum: Slit B Absent}

The inner curve of Figure 5 of \cite{Shih1} is drawn from the experimental coincidence counts obtained
when slit B is ``wide open''.  This shows that the photons are deflected
up to 0.9 mm, and that from 0.9 mm to 1.45 mm the detection rate is constant and close to zero.

\subsubsection*{Interpretation as Diffraction from Ghost slit B}   
If this inner curve is interpreted in the same way as the outer curve (as in the immediately preceeding 
section), then the ``first diffraction minimum'' would be at $y = 0.9$ mm.
The time taken for the photon
to travel from the ghost slit B to the detector D2 is the same as when slit B
 is actually present; i.e.~$0.500/c=1.66{\scriptstyle 78}\times 10^{-9}$ seconds.  

In this time it travels 0.9 mm in the $y$ direction; i.e.~its speed in the $y$ direction is: 
\begin{equation}\label{vy}
v_y = 0.0009/1.66{\scriptstyle 78}\times 10^{-9} = 5.39{\scriptstyle 63}\times 10^{5} {\rm metre/sec}
\end{equation}
and hence its transverse momentum (in the $y$ direction) is:
\begin{eqnarray}\nonumber
P_y = M\,v_y = 3.147{\scriptstyle 61}\times 10^{-36}\times 5.39{\scriptstyle 63}\times 10^{5} 
&=& 1.69{\scriptstyle 85}\times 10^{-30} \;{\rm Kg.metre/sec}
\end{eqnarray}
Multiplication of this transverse momentum by the slit width of 0.16 mm produces a position-momentum
uncertainty product of:
\begin{eqnarray}\nonumber
\Delta y\,\Delta P_y 
&=& 0.00016\times 1.69{\scriptstyle 85}\times 10^{-30}
\\
&=& 2.7{\scriptstyle 18}\times 10^{-34} \;{\rm Joule.sec}
\end{eqnarray}
which is smaller than Planck's constant:
\begin{eqnarray}\nonumber
h &=& 6.626176\times 10^{-34} \;{\rm Joule.sec}
\end{eqnarray}
the ratio being:
\begin{eqnarray}\nonumber
2.7{\scriptstyle 18}\times 10^{-34}/6.626176\times 10^{-34} = 0.41
\end{eqnarray}
an apparent violation of the uncertainty principle as formulated by Feynman; i.e. a violation
of eqn.(\ref{HUPdiffdelta}).  For this reason this interpretation must be wrong.  

The result vindicates Popper  when he wrote that he was \cite[p.29]{Popper}:
\begin{quote}
``inclined to predict''  that in an actual experimental test, the photon at
the location of the absent slit B would not diffract (``scatter'') when its partner photon (with which it
is in an entangled state) is localized within slit A.  
\end{quote}
It is also a vindication of Einstein, Podolski, and Rosen, in their inference that such 
instantaneous\footnote{or at least superluminal with no upper limit on the speed of transmission}
action at a distance is inconsistent with the causality principle of Special Relativity -- that physical effects
cannot travel faster than the speed of light.  Since this experimental result is consistent with causality,
the further inference of EPR (that quantum theory must be incomplete) is also vindicated. 

\subsubsection*{Interpretation as Diffraction from the Source}   

In this case the photons have
traveled a distance of 1245 mm from their source (without encountering any slit).
The time taken for the photon
to travel 1245 mm along the $x$ axis is $1.245/c=4.152{\scriptstyle 87}\times 10^{-9}$ seconds.  
In this time it travels 0.9 mm in the $y$ direction to the ``first minimum'' in the diffraction pattern; 
i.e.~its speed in the $y$ direction
is $0.0009/4.152{\scriptstyle 87}\times 10^{-9} = 2.16{\scriptstyle 72}\times 10^{5}$ metre/sec, 
and hence its transverse momentum is:
\begin{eqnarray}\nonumber
P_y = M\,v_y = 3.147{\scriptstyle 61}\times 10^{-36}\times 2.16{\scriptstyle 72}\times 10^{5}
&=& 6.821{\scriptstyle 42}\times 10^{-31} \;{\rm Kg.metre/sec}
\end{eqnarray}
Multiplication of this transverse momentum by the slit width of 0.16 mm produces a position-momentum
uncertainty product of:\footnote{This multiplication by a slit width of 0.16 mm is dubious because the
diffraction is assumed to originate in the source (the BBO crystal) rather than at the 
ghost image of slit A.}
\begin{eqnarray}\nonumber
\Delta y\,\Delta P_y 
&=& 0.00016\times 6.821{\scriptstyle 42}\times 10^{-31}
\\
&=& 1.091{\scriptstyle 42}\times 10^{-34} \;{\rm Joule.sec}
\end{eqnarray}
which is smaller than Planck's constant, the ratio being:
\begin{eqnarray}\nonumber
1.091{\scriptstyle 42}\times 10^{-34}/6.626176\times 10^{-34} = 0.1647
\end{eqnarray}
which is an even greater violation of the Uncertainty Principle (of eqn.(\ref{HUPdiffdelta})) than
when interpreted as diffraction from the ghost slit B.

\subsubsection*{Non-Diffractive Interpretation}

The inner curve of Figure 5 of \cite{Shih1} does not display a discernable diffraction 
minimum.\footnote{unlike the wider curve}  There are in fact 5 data points 
($y= 1.0 - 1.45$ mm) all of which have the same (very small) value; this suggests that the 
origin of this peak is {\em not} diffraction through a slit.  

\paragraph{It is noteworthy that Kim and Shih report}\cite[p.1856,$2^{\rm nd}\P$]{Shih1}
\begin{quote}
``the single detector counting rate of D2 keeps \\\ \ \ constant in the entire scanning range''
\end{quote}
In other words: when the coincidence circuit is switched off  D2 detects the same count rate at
all values of $y$ at which it was placed (presumable from $y=0$ to $y=1.45$ mm); this would be
a horizontal straight line if added to Figure 5 of \cite{Shih1}.
This measurement was simply seeing the beam from the source towards D2 
with a uniform intensity over the scanning range of $\approx$ 3 mm ($y=\pm 1.5$ mm).

\hspace*{0.25in}Kim and Shih also report \cite[p.1856,$2^{\rm nd}\P$]{Shih1}:
\begin{quote}
``the width of the pattern is found to be much narrower \\\ \ \ than the actual size of the diverging SPDC 
beam at D2''
\end{quote}
The alternative, plausible cause of this narrow peak is a convolution of the finite size of the 
source,\footnote{a cylinder 3 mm diameter and 3 mm long} with the geometry of possible
coincidences.  

\hspace*{0.25in}Further clarification on the origin of this peak depends upon experimental details which 
are not available in the publications; in this regard it is a pity that raw data 
(such as actual, observed counting rates) were not included in the published reports of the experiment.

\section{A Retrodictive Realist Account}
\begin{itemize}
\item A pair of photons produced by SPDC is in an entangled state from the moment of generation 
 until one of them enters a slit; entanglement means that their positions
and momenta are correlated; knowledge of the position of one photon allows one to infer the
position of the other photon; likewise for their momenta.
\item When one photon enters a slit it interacts with that slit and this destroys the coherence
(entanglement) between them; i.e.~decoherence occurs.  
The interaction can be attributed to the photon being some kind of
localized electromagnetic wave, which interacts with the electrons in the surface of the solid
that forms the slit.  That photons are localized waves is supported experimentally by the production
of laser pulses as short (in time) as two optical periods; thus the photon cannot be longer than
two wavelengths along its direction of propagation.
\item Measurement of the diffracted position ($y$ coordinate) of a photon 
(coincidence detection by D2) with slit B absent, allows one to calculate not only the momentum
vector of this photon as in travels from the source to D2, but also the momentum of the other
photon as it travels from the source to slit A; however when this latter photon enters slit A its
interaction with the walls of the slit causes it to diffract at an angle which is predictable only
statistically - in accord with the uncertainty principle.  

\item Thus coincidence measurements \underline{with slit B absent} provide the positions of both photons
(from the detection of a photon having passed through slit A) with a precision equal to the
width of slit A.  Likewise the measurement of the deflected position ($y$ coordinate) of a photon by D2
allows one to calculate the momentum vectors of {\em both} photons of the entangled pair -- during their
trajectories from the source to the plane of slit A (for one photon), and from the source
to the scanning plane of D2 for the other photon.  These {\em in principle}, precise, retrodictive 
calculations of the trajectories of both photons are unfortunately limited in precision by the actual 
experimental results because of the relatively large size of the non-point source.\footnote{a cylinder 
3 mm diameter and 3 mm long}

\item It is especially noteworthy that \underline{individual events}\footnote{the generation and detection
of a particular entangled photon pair} are not limited by the uncertainty principle: any diffraction of
a photon to a position of D2 smaller than the $y$-coordinate of the first diffraction minimum will
yield a position-momentum product that is smaller than Planck's constant -- even when slit B 
is present.  In particular, the most probable diffraction angle (to the top of the central peak)
yields a transverse momentum of zero, which when multiplied by the uncertainty in its
position (the slit width of 0.16 mm) yields an uncertainty product of zero !
\end{itemize}

This realist interpretation is in accord with the {\sl consistent histories} interpretation of 
quantum mechanics \cite{OmnesBook,Omnes}; as the term ``history'' implies, it as a retrodictive
realism.   However, experimental physics is largely concerned with retrodictively interpreting
the results of measurements; hence as Feynman \cite{Feynman} and Holland \cite{Holland}
have noted, quantum mechanics does not preclude the precise {\em retrodictive} description of past
events in terms of the classical coordinates and momenta of particles; it only precludes predictions.

\hspace*{0.25in}The long standing controversy over ``hidden variables'' \cite{Belinfante} must now 
be seen as a ``red herring'', for the
``hidden variables'' in this realization of Popper's experiment are the precise location of the source of an 
individual photon pair within the BBO crystal, and the impact parameter of each photon with the slit
that it passes through.  These ``hidden variables'' would more accurately be described as
``uncontrollable parameters'' of an individual two-photon event; an essential, statistical uncertainty
arising from the experimental arrangement.

\hspace*{0.25in}It is especially noteworthy that this interpretation -- that each photon has a classical
trajectory except when it is interacting with a slit or detector -- is inconsistent with the Bohmian
interpretation of quantum mechanics \cite{Holland}, because there is no 
``quantum potential'' present to affect the classical trajectories.

\section{The Incompleteness of Quantum Mechanics}
Notwithstanding developments in the formalism of quantum mechanics \cite{Rae,Kaempffer}:
\begin{itemize}
\item the formulation in terms of {\sl Positive, Operator-Valued, Measures} \cite{Nielsen,Peres} and 
\item the formulation as a strictly operationist theory \cite{Busch},
\end{itemize}
the ``measurement problem'' remains problematical and essentially 
unresolved\\ \ \cite[Ch.V,pp.131-137]{Busch2}.  
Holland \cite[pp.328-333]{Holland} has inferred that
\begin{quote}
``If we cannot account for the measuring process by applying the usual\\\ \ many-body 
Schr\"odinger theory,
this implies a massive incompleteness\\\ \ in the quantum mechanical treatment of general natural 
processes''.
\end{quote}
Thus regardless of the question of realism (and of experimental results in accord with
 realism or otherwise)
quantum theory has been recognized to be logically incomplete 
because it does not describe the actual physical processes involved in making measurements. 
 It is also incomplete because it does not describe individual events, whereas individual events
are commonly observed in experimental physics.

\subsection*{Concluding Remarks}

The above assertion of a classical, retrodictive and realist interpretation of the realized 
Popper's experiment, is not easily extended to the interpretation of some other experiments, notably the
various double path experiments that have been conducted.  

\hspace*{.25in}A plausible, yet tentative, realist
interpretation of how a molecule as large as a ``bucky ball'' (C$_{60}$) can ``go through 2 slits
at the same time and interfere with itself'' \cite{Zeilinger1} is that every particle is surrounded by
a real oscillating field, which manifests itself in the experiments as a wave with an effective
wavelength given by the de Broglie relation in terms of the particle's laboratory-frame momentum.

\hspace*{.25in}However, double-path experiments of the Mach-Zender type\footnote{as distinct from 
double-slit experiments} defy any realist interpretation because the two paths are separated by a
macroscopic distance thus making the idea that the particle goes along one path while its
surrounding wave-field interacts with the other path, implausible; this is especially so when the 
particle that goes along
both paths and interferes with itself is electrically neutral; e.g.~the neutron interferometry experiments
of Zeilinger et al \cite{Zeilinger2}.  This phenomenon of the self-interference of a wave propagating
along widely separated paths is a form of wierdness regardless of quantum mechanics; its most
bizarre manifestation is in long-baseline interferometry in radio astronomy, where the two
paths are separated by many thousands of miles.

\hspace*{.25in}
While Popper was correct in regarding his experiment as a ``crucial'' test of the inconsistency between:
\begin{itemize}
\item the presumed non-locality of quantum mechanics manifest by instantaneous actions at a 
distance, and
\item the causality principle of Special Relativity
\end{itemize}
and while this realization of Popper's experiment supports the conclusion that the inferences of such   
instantaneous actions at a distance are the result of incorrect quantum theoretical 
argument,\footnote{because the experiment results do not display the expected actions at a distance}
it is nevertheless salutory to recognize that this experimental disproof of instantaneous actions at a 
distance, does little to resolve the dilemma of double-path interference in locally physical terms.

\hspace*{.25in}The zeal with which the completeness of quantum mechanics was defended by Bohr,
Heisenberg, and their like-minded peers, can be understood as the enthusiasm of the proponents
of a new\footnote{new in the 1920s and 1930s} and manifestly successful theory\footnote{Its early
successes were the calculations on the hydrogen atom (Schr\"odinger, 1926), quickly followed by
accurate calculations on the helium atom, and the molecules H$_2^+$ and $H2$.} which they felt
impelled to defend against the skepticism of such distinguished elder physicists as Eddington 
\cite[Ch.X,pp.200-229,p.222]{Eddington}.  It is salutory to reflect that the papers and letters of the
1930s were written in the midst of the battle to understand and establish quantum mechanics.

\hspace*{.25in}
It is pity that this justifiable enthusiam led them to claim more\footnote{that it was a complete theory
of the physical world} than the theory's ambit of application warranted.  The concurrence of the
majority of physicists with this undue claim of completeness is understandable, because tacit 
acceptance of something you don't really understand\footnote{recall Feynman's edict that ``nobody 
understands quantum mechanics''.} 
is easier than engaging in an intellectual struggle for understanding especially when the topic
doesn't seem to have much relevance to ones daily work.  Recall once again Popper's observation:
\begin{quote}
``my assertion [is] that most physicists who honestly believe in the \\\ \ Copenhagen interpretation
do not pay any attention to it in actual practice''.
\end{quote}
This acquiessence in the Copenhagen interpretation is common societal phenomenon arising from the
predisposition of a person to agree with his/her companions; it was satired in
the story of the courtiers' mutual agreement about the splendor 
of the Emperor's New Clothes, when in fact the Emperor was naked.

\hspace*{.25in}
It is ironic that this uncompromising (albeit tacit) advocacy of the completeness of quantum 
mechanics has now made its advocates the conservative defenders of a proposition that nobody really
understands.  The statements by Kim and Shih \cite[pp.1858-1859]{Shih1} in defense of 
quantum theory seem to be motivated by just such an irrational faith.\footnote{or perhaps they
were included to circumvent, or in response to, a negative evaluation by the journal's referees} 
Thus as other philosophers have noted, youthful rebels turn into conservative bigots in their later years. 
\begin{center}
{\sl fiat lux}
\end{center}

\section*{Acknowledgements}

The support of the Natural Sciences and Engineering Research Council of Canada is gratefully
acknowledged.  

\hspace*{.25in}
Thomas Angelides brought the Shih-Kim experiment to the
author's attention at conferences in Baltimore (1999) and Berkeley (2000).  

\hspace*{.25in}
Yanhua Shih provided helpful elaborations of the published accounts of the experimental work and its
interpretation (sent to the author by email).

\hspace*{.25in}
John Sipe of the University of Toronto is thanked for 
allowing the author to participate in his course, ``Current Interpretations of Quantum Mechanics'',
given January-April 2002; this stimulated the author's interest which led to the writing
of this article.  

\hspace*{.25in}
Yasaman Soudagar is thanked for bringing John Sipe's innovative course to the author's
attention.  

\hspace*{.25in}
Marian Kowalski is thanked for many congenial discussions with a kindred spirit believing
(like Einstein, Popper, Schr\"odinger, et al) in the essential simplicity and logicality of individual
events in the physical world notwithstanding the limitations on predictive measurements arising from the
finite value of Planck's constant and the inevitably uncontrollable parameters of each experimental
arrangement.

\end{document}